\documentclass[10pt,letterpaper,twocolumn]{article} 
\usepackage{ol2}

\usepackage[utf8]{inputenc}
\usepackage{graphicx}

\usepackage{hyperref}
\usepackage{amsmath}
\usepackage{amssymb}
\usepackage{amsfonts}
\usepackage{latexsym}
\usepackage{pifont}

\begin{document}

\twocolumn[ 

\title{Fiber-assisted single-photon spectrograph}
\date{\today}

\author{Malte Avenhaus$^*$, Andreas Eckstein, Peter J.~Mosley, and Christine Silberhorn}
\affiliation{Max Planck Institute for the Science of Light, Junior Research Group IQO \\ G\"unther-Scharowsky-Stra\ss{}e 1/Bau 24, 91058 Erlangen, Germany \\
$^*$Corresponding author: malte.avenhaus@mpl.mpg.de}

\begin{abstract}
We demonstrate the implementation of a fiber-integrated spectrograph
utilizing chromatic group velocity dispersion (GVD) in a single mode fiber. By means of GVD we stretch an ultrafast pulse in time in order to spectrally resolve single photons in the time domain, detected by single photon counting modules with very accurate temporal resolution. As a result, the spectrum of a very weak pulse is recovered from a precise time measurement with high signal to noise ratio. We demonstrate the potential of our technique by applying our scheme to analyzing the joint spectral intensity distribution of a parametric downconversion source at telecommunication wavelength.
\end{abstract}


] 

\label{sec:motivation}

The success or failure of many experiments in quantum optics such as conditional quantum state preparation \cite{Lvovsky,Dakna} or two photon interference depends critically on the intrinsic spatio-spectral structure \cite{Rohde:2005p7472} 
of the light used. Recently it has been shown that the use of ultrafast pump pulses for generating non-classical light by a non-linear process  often leads to a complex multi-mode structure \cite{Law,Opatrny,Wasilewsky1,deValcarcel} of quantum states. This is strongly related to spectral correlations amongst the different frequency components of the created photon pairs. The degree of spectral entanglement between a pair of photons is of particular importance since it imposes on the one hand a severe limitation upon the purity of heralded single photon Fock states \cite{URen:2005p7251}, and on the other hand it can serve as a resource for quantum communication applications. Therefore several groups have recently started to control \cite{Branning,Walton,Valencia,Mosley}  
and characterize \cite{Kim,Wasilewsky2,Poh,Baek1,Kuzucu} 
the bi-photon joint correlation spectra of parametric downconversion (PDC) sources. 

In classical optics there exist several methods of spatially separating individual frequency components of a light field \cite{Mandel, Saleh}.  Most types of spectrometers can be subdivided into two classes:  scanning devices and spectrographs. Scanning spectrometers like monochromators generally employ moving parts, but they are comparatively cheap to produce for they only require a single detector. Yet, this comes at the expense of discarding information since they record only one spectral component at a time. Spectrographs typically consist of detector arrays like CCDs to determine all spectral components simultaneously. While this approach is feasible for bright light it is too noisy and expensive for single photons. The detection technique that is conventionally used for single photon sensitivity employs an internal gain mechanism that makes use of an avalanche process, as in avalanche photo diodes (APD).

To obtain detailed knowledge about the shape of an optical spectrum proves to be extremely challenging on the single photon level. The necessity of retaining high sensitivity at very low loss greatly reduces the practicability of conventional spectrometers based on gratings, prisms or etalons.
Here, we present a cost effective solution to this problem that enables the precise and efficient measurement of single photon spectra for ultrafast light pulses whilst maintaining an excellent signal to noise ratio and easy alignment. It has been shown previously \cite{Valencia2,Brida,Baek}, but not developed as a measurement technique, that a single photon becomes chirped due to group velocity dispersion (GVD) in a fiber. Our method exploits the much higher absolute value of GVD available in dispersion compensating fibers (DCF) to stretch a short pulse such that different spectral components map to different arrival times at the detector. As a single detector is sufficient to monitor all possible  arrival times, one can record the complete spectrum of a pulse without the need for a multi-element detector array. This is particularly apposite when measuring a pulse at telecommunication wavelengths where detector arrays are very expensive and, in the low photon number case, prohibitively noisy. Hence, as only a single APD is required, our scheme can in principle combine the informational advantage of a spectrograph with the economical advantage of a scanning spectrometer. We demonstrate the applicability of our scheme experimentally by measuring the single photon marginal as well as joint correlation spectrum of a PDC source at a wavelength of around 1550\,nm.

\label{sec:calibration}

\begin{figure}[bp]
\includegraphics[width=\linewidth]{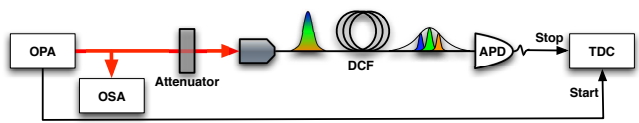}
    \caption{Calibration setup: A reference spectrum is coupled into a DCF (see text). An APD detects the photon's arrival time which is subsequently recorded with a TDC.}
    \label{fig:CalibrationSetup}
\end{figure}

In the first step, we used the experimental setup depicted in Fig.~\ref{fig:CalibrationSetup} to calibrate and to investigate the achievable resolution of the DCF spectrometer, by measuring the chromatic GVD in our fiber. A calibration procedure needs to be performed against a trusted reference spectrometer: here, this refers to mapping the travel time through the fiber to a specific wavelength. We utilized a pulsed and tunable optical parametric amplifier (OPA) system as the light source and monitored the central wavelength $\lambda_C$ with an optical spectrum analyzer (OSA). The beam was strongly attenuated to yield on average less than one photon per pulse before being coupled into a \emph{DK-40} dispersion compensating fiber module from Lucent. The photons were then detected by an \emph{id201} InGaAs APD from id-Quantique with a measured temporal jitter of only $\sigma_{\tau, \mathrm{APD}}$=180\,ps. 
We recorded the time difference $\Delta \tau$ between the electronic laser trigger and the electronic response from the APD with a time-to-digital converter (TDC) for a set of three different spectra with $\lambda_C=$1531\,nm, 1484\,nm and 1391\,nm, and observed corresponding peaks at $\Delta \tau=$1874\,ns, 1884\,ns and 1990\,ns. Utilizing the TDC, we were able to quantify $\Delta \tau$ with a precision of 81\,ps.

\begin{figure}[bp]
\includegraphics[width=\linewidth]{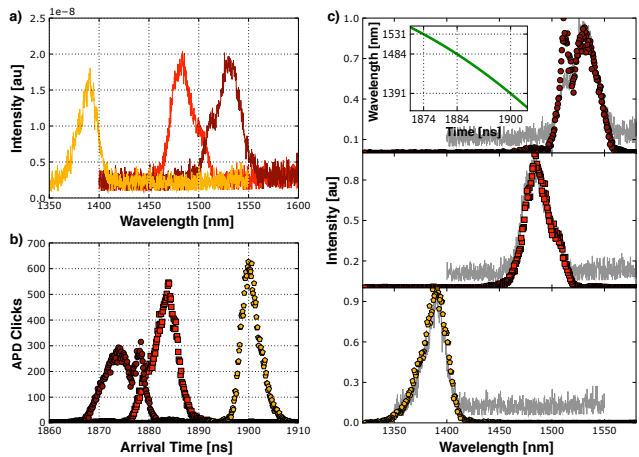}
    \caption{Calibration using three reference spectra: a) OSA reference spectrum, b) Arrival time $\Delta \tau$, c) Comparison of original and reconstructed spectra.}
    \label{fig:Calibration}
\end{figure}

By relating the peaks from the OSA spectrum with the temporal peaks from the time measurement (see Fig.~\ref{fig:Calibration}) we obtained a calibration curve (inset) for reconstructing the entire set of spectra. A least square polynomial fit relating the arrival time $\tau$ to the monitored wavelength was used as the calibration curve $c(\tau)$. It is clearly evident from Fig.~\ref{fig:Calibration}c that the spectral width and shape of the OSA measurement overlap well with the reconstructed spectra. Although the input pulse is strongly attenuated, the DCF measurements show much better signal-to-noise ratio, and the side lobe of the darkest colored curve is much more pronounced, indicating an excellent resolution.

Furthermore, we investigated the capabilities, potential limitations and experimental constraints encountered when using our setup for recovering a spectrum. The derivative of the quadratic fit function gives a GVD between -0.11\,ns/nm at 1325\,nm and -0.25\,ns/nm at 1575\,nm. With a detection jitter of $\sigma_{\tau, \mathrm{APD}}=$180\,ps, a resolution of 0.72\,nm can be achieved at 1550\,nm. Resolution can be increased using a longer fiber at the cost of higher losses and greater expense. The fit for the calibration function $c(\tau)$ to three calibration wavelengths is an approximation that may give rise to systematic errors. We estimate these errors to be of magnitude $\pm$0.1\,nm at around 1520\,nm and $\pm$0.5\,nm at around 1550\,nm. This uncertainty poses no principal restriction and can be reduced by using more reference spectra for the calibration. If the laser repetition rate is above a certain threshold it is possible that several light pulses are present simultaneously in the DCF. This will only cause problems for the reconstruction if the spectrum of a pulse is so broad that it becomes mixed with the following pulse. We avoided this problem by reducing the 80MHz repetition rate of our laser down to 1MHz with a pulse picker. We determined the travel time through the fiber to be 16.6\,$\mu$s, or equivalently a fiber length of 3.3\,km. In addition, APDs exhibit a wavelength dependent variation of the single photon detection probability $p_D(\lambda)$. Hence, it is crucial to renormalize the recorded click histogram against the detection probability $p_D(\lambda)$ if the wavelength range under observation is too large for $p_D(\lambda)$ to be assumed flat.


\label{sec:parametric_down_conversion}
\begin{figure}[bp]
\includegraphics[width=\linewidth]{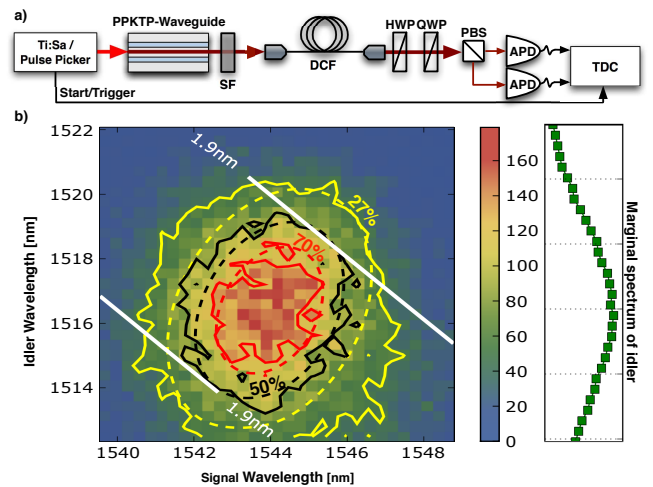}
  \caption{a) Experimental setup for measuring marginal and joint correlation spectra of a PDC source (see text). Not shown is the programmable delay generator used for gating the APDs.
  b) Measured correlation spectrum between signal and idler corresponds to the joint spectral intensity $|F(\omega_s, \omega_i)|^2$. A white line marks the 1.9nm FWHM of the pump envelope. Solid and dashed contour lines are shown for theory and experiment respectively.}
  \label{fig:jsi}
\end{figure}

After having ensured a reliable reconstruction of single photon spectra we employed our scheme to
measure the spectrum of a PDC source. Our setup for this application of the spectrograph is shown in
Fig.~\ref{fig:jsi}a and allowed us to analyze the marginal as well as joint correlation spectrum of
the PDC. We pumped an 18\,mm long PPKTP waveguide with a $4\mu\text{m} \times 4\mu\text{m}$ cross
section and generated a wavefunction of two photons
\[ |\Psi\rangle = \int \text{d}\omega_s \text{d}\omega_i \;F(\omega_s,  \omega_i)\;a^\dagger(\omega_s)b^\dagger(\omega_i) |0,0\rangle \]
with a spectral amplitude function $F(\omega_s, \omega_i)$. Our goal was to analyze precisely the
bi-photonic correlation properties of the spectral intensity distribution$|F(\omega_s, \omega_i)|^2$.

The PDC process was pumped by an ultrafast pulsed Ti:Sa laser system at 765\,nm with FWHM of 1.9\,nm to produce photon pairs at 1544\,nm and 1517\,nm in a type-II process, such that signal and idler photons emerged from the waveguide with orthogonal linear polarization. Both photons were coupled into the DCF with 25\% efficiency. Since the fiber was not polarization preserving, signal and idler photons carried elliptic, but still orthogonal, polarization states at the fiber output. To restore their linear polarization we applied a quarter- and half-waveplate. Finally, a PBS split signal and idler, and they were guided to \emph{id201} InGaAs APDs. 

As the polarization rotation caused by the DCF was wavelength dependent we could not perfectly reconstruct the original linear polarization and thus unambiguously split signal and idler photons for their entire spectrum. Nevertheless, we were able to achieve a reasonable polarization contrast of 80\% for both signal and idler beams. Since signal and idler were not degenerate in wavelength, we could identify and discard the photons that took the 'wrong' path at the PBS by their arrival time, and thus the quality of our measured spectra were not degraded. In principle, all polarization-related problems would be avoided by separating signal and idler immediately after the waveguide. This, however, requires two DCFs.

In the telecommunication wavelength regime, due to the small semiconductor bandgap, current APDs need to be operated in a gated mode to suppress dark counts and afterpulsing. We thus scanned the time domain by electronically delaying the laser trigger in steps of 100\,ps. We selected the shortest possible gate width such that all detection events occurred in a $\sigma_{\tau, \mathrm{APD}}\approx$200\,ps window. The delay was independently varied for signal and idler. Each coincident detection event yielded a pair of arrival times that could be mapped to wavelengths with the measured calibration curve via $(\lambda_s, \lambda_i) = (c(\tau_s), c(\tau_i))$.

The calibration curve $\lambda = c(\tau + \delta\tau)$ is unique up to an offset $\delta\tau$ which is introduced when the spectrometer is used in a setting where the optical path length differs from that of the calibration setup. We derived this offset by taking into account energy conservation 
$\frac{1}{\lambda_p}=\frac{1}{\lambda_s}+\frac{1}{\lambda_i} = \frac{1}{c(\tau_s+\delta \tau)}+\frac{1}{c(\tau_i+\delta \tau)}$
of the PDC process. With this offset we were able to reconstruct both the marginal and the coincidence spectrum of our PDC source with high precision, as shown in Fig.~\ref{fig:jsi}b.

\label{sec:conclusion}
In summary, we have introduced a new technique for measuring directly the spectrum of ultrafast pulses at the single photon level using GVD in a highly dispersive fiber. We showed how to calibrate and apply such an apparatus for measurements at wavelengths of $\approx$1550\,nm with both high resolution and high signal-to-noise ratio. Thereafter, we enhanced our scheme to measure two-dimensional coincidence spectra, and thus demonstrated a characterization of the joint spectral intensity of signal and idler photons from a PDC source at telecommunication wavelengths. Our approach might become even more attractive for current experiments performed at 800\,nm as this offers the advantage of making full use of the DCF as a spectrograph, since no gating is required by visible light sensitive Si-APDs. 
Because of the simplicity and accuracy of our experimental setup we expect the DCF spectrograph to become a versatile tool for the characterization of optical quantum states.


\label{sec:so_long_and_thanks_for_all_the_fish}

We thank Bernhard Schmauß for his support and providing the DCF. We acknowledge the financial support by the European Commission from the Future and Emerging Technologies (FET) project CORNER 
under contract FP7-ICT-213681.


\end{document}